\documentclass[12pt]{iopart}

\usepackage{graphicx}
\usepackage{xfrac}
\usepackage{epsfig}
\usepackage{iopams}
\usepackage{color}

\newcommand{\dc}{\textsc{dc}}
\newcommand{\ac}{\textsc{ac}}
\newcommand{\jrm}{\textsc{jrm}}

\begin{document}
\title[Transmon Quantum Annealer]{A Transmon Quantum Annealer: \\
Decomposing Many-Body Ising Constraints Into Pair Interactions}

\author{ Martin Leib$^{1,2}$, Peter Zoller$^{1,2}$, Wolfgang Lechner$^{1,2}$}
\address{$^1$Institute for Quantum Optics and
Quantum Information, Austrian Academy of Sciences, 6020 Innsbruck, Austria}
\address{$^2$Institute for Theoretical Physics,  University of Innsbruck,
6020 Innsbruck, Austria}

\date{\today}


\begin{abstract} 

Adiabatic quantum computing is an analog quantum computing scheme with various applications in  solving optimization problems. In the parity picture of quantum optimization, the problem is encoded in local fields that act on qubits which are connected via local 4-body terms. We present an implementation of a parity annealer with Transmon qubits with a specifically tailored Ising interaction from Josephson ring modulators. 
\end{abstract}

\submitto{\NJP}

\maketitle
\tableofcontents

\section{Introduction}

Among superconducting qubits, Transmons are a leading platform with respect to decay and dephasing times \cite{WALLRAFF,KOCH,SCHOELLKOPF,MARTINIS}. They are charge qubits operated in a regime with strong resilience to the ubiquitous charge noise in any superconducting qubit devices. With this insensitivity to radio frequencies and tuneability by fast microwave signals they are considered ideal candidates for gate base quantum computing. For analog applications such as adiabatic quantum computing, Transmons have not been considered so far.
\begin{figure}[htb]
\centerline{\includegraphics[width=7cm]{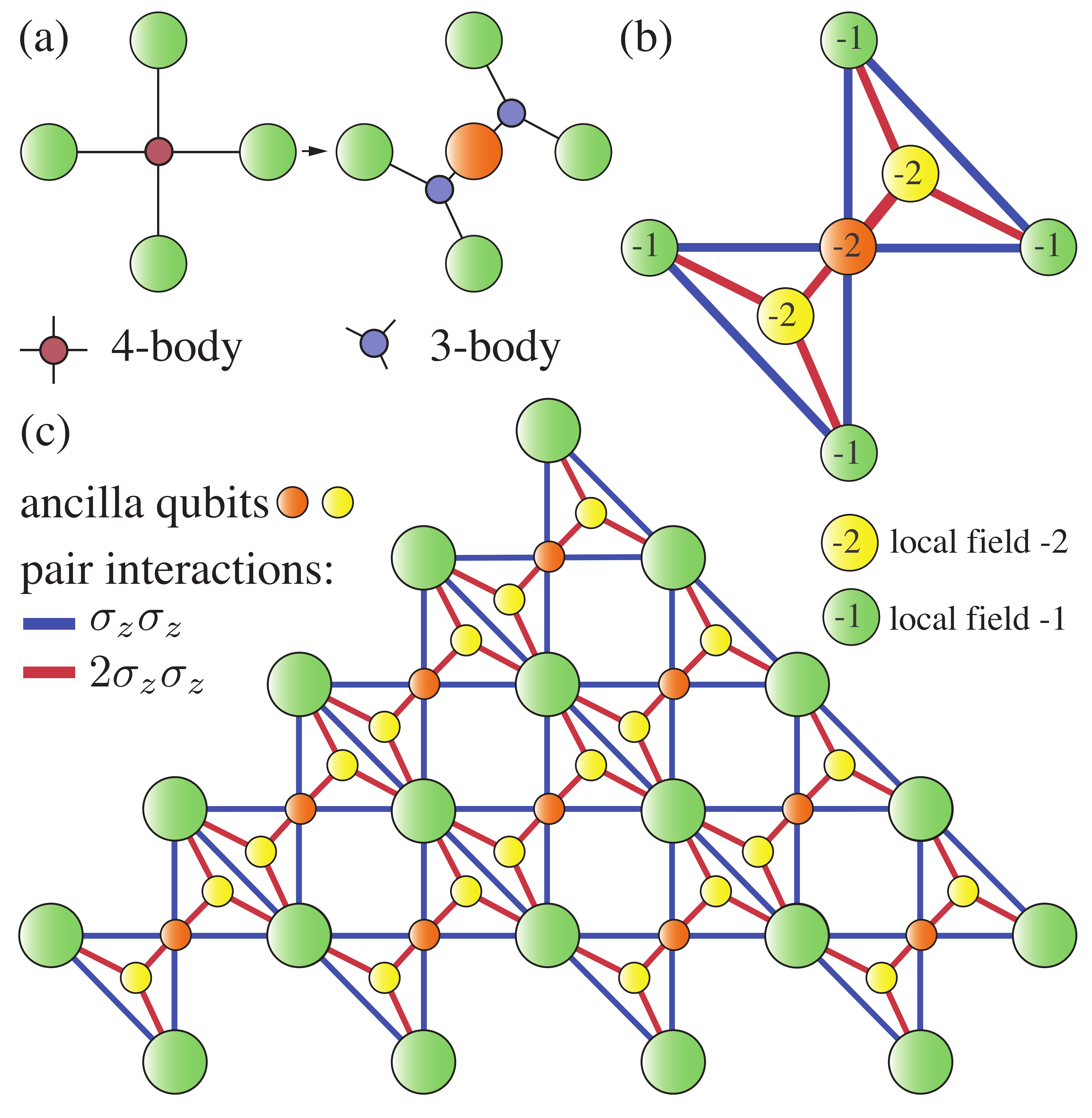}}
\caption{(a) The 4-body constraints (red) of the parity quantum annealing scheme are decomposed into two 3-body terms (blue) with one ancilla qubit (orange). (b) In a second step, the 3-body terms are decomposed into pair interactions with an additional ancilla (yellow). The interaction strengths are either $+1 \sigma_z\sigma_z$ (blue) or $+2 \sigma_z\sigma_z$ (red). Local fields acting on ancilla qubits and programmable qubits are $-2$ and $-1$, respectively. (c) The architecture based on these plaquettes consists of $K=N(N-1)/2$ programmable qubits and  $M=(N-2)^2/2$ ancilla qubits.}
\label{fig:illustration}
\end{figure}

Adiabatic quantum computing \cite{FAHRI} has recently gained considerable attention because of the prospect to solve problems that can be mapped to classical optimizations \cite{CLASSIFIER,CHEMISTRY,SEARCH,PROTEIN}  efficiently \cite{SCIENCE_QA,Boixo2013,Boixo2014,HARRIS,JOHNSON,Steffen2003,SCIENCE_SPEEDUP}. In the spin-glass formulation of adiabatic quantum annealing \cite{FAHRI}, the system is slowly switched from the ground state of a trivial initial Hamiltonian in $\sigma_x$-basis to an infinite range Ising model in $\sigma_z$-basis \cite{LUCAS}. The optimization problem is encoded in the interactions between the spins in the final state. There are major challenges related to this protocol: A fundamental challenges in this scheme is the required all-to-all connectivity \cite{KATZGRABER,SPINGLASS}. Infinite range connectivity are required while natural qubit interactions are finite range \cite{EMBEDDING,EMBEDDING2}. Another fundamental question is the possible quantum speedup due to the scaling of the minimal gap \cite{SCIENCE_SPEEDUP} and the sensitivity to errors \cite{LIDAR}. A proposal with the aim to address several of these challenges is the parity adiabatic quantum optimization scheme \cite{PROPOSAL}. In this model, a fully connected system is encoded in a larger Hilbert space with local constraints on a square lattice. The optimization problem is, in contrast to embedding schemes\cite{EMBEDDING,EMBEDDING2}, encoded in local fields that act on the qubits and the interactions are 4-body constraints that are independent of the problem. 

In this paper we propose an implementation of the parity adiabatic quantum optimization scheme \cite{PROPOSAL} with pair with Transmons. For this purpose, we present a general recursive decomposition of classical k-local Ising terms. Unlike previous proposals for the realization of such many-body terms\cite{SOLANO}, in our work, the higher order terms are not the result of a perturbation theory, a time-dependent model or a gadget construction. Rather, we focus on only reproducing the ground state which allows one to decompose the constraints to direct pair interactions. Our construction consists of two steps: a recursive decomposition of k-local terms to 3-body terms and a decomposition of each 3-body term into two-body terms. Based on this decomposition, we present a Transmon implementation with a 2D setup where all pair interactions have the same sign, are next neighbor and do not cross each other. In the parity scheme, the $\sigma_z\sigma_z$ interactions can be always-on which enables an implementation with coupled Josephson ring modulators. 

The paper is structured as follows: In Section 2.1 we briefly review the parity adiabatic quantum computing scheme. In Section 2.2 we present the general decomposition of Ising-type constraints which we apply in Section 2.3 to the 4-body terms of the PAQC scheme.

\section{Parity Adiabatic Quantum Computing with pair interactions}
AQC maps the solution to a complicated optimisation problem into the groundstate of a specifically designed Hamilton operator $H^{(\text{final})}$ \cite{LUCAS}. Rather than cooling into the groundstate of a system that is described by $H^{(\text{final})}$, which is two slow, one prepares the system at a different point in  parameter space described by $H^{(\text{initial})}$ in its groundstate. Afterwards a adiabatic sweep in parameter-space prepares the system in the groundstate of the problem Hamilton operator $H^{(\text{final})}$. This is only advantageous if there is a known efficient technique to prepare the system in the groundstate of $H^{(\text{initial})}$ and if the maximal sweep speed scales polynomial with the problem size. One can think of AQC as the slow deformation of a complex high dimensional energy landscape. The state of the system is prepared in a trivial valley of the energy landscape and the adiabatic sweep deforms the energy landscape while the system tries to stay in the instantaneous valley. To unlock the true potential of quantum mechanics we have to enable the system to tunnel through energy barriers to the deepest valley. This is accomplished by choosing initial and final Hamilton operators that do not commute  $[H^{(\text{final})},H^{(\text{initial})}]\neq 0$

\begin{figure}[htb]
\centerline{\includegraphics[width=16cm]{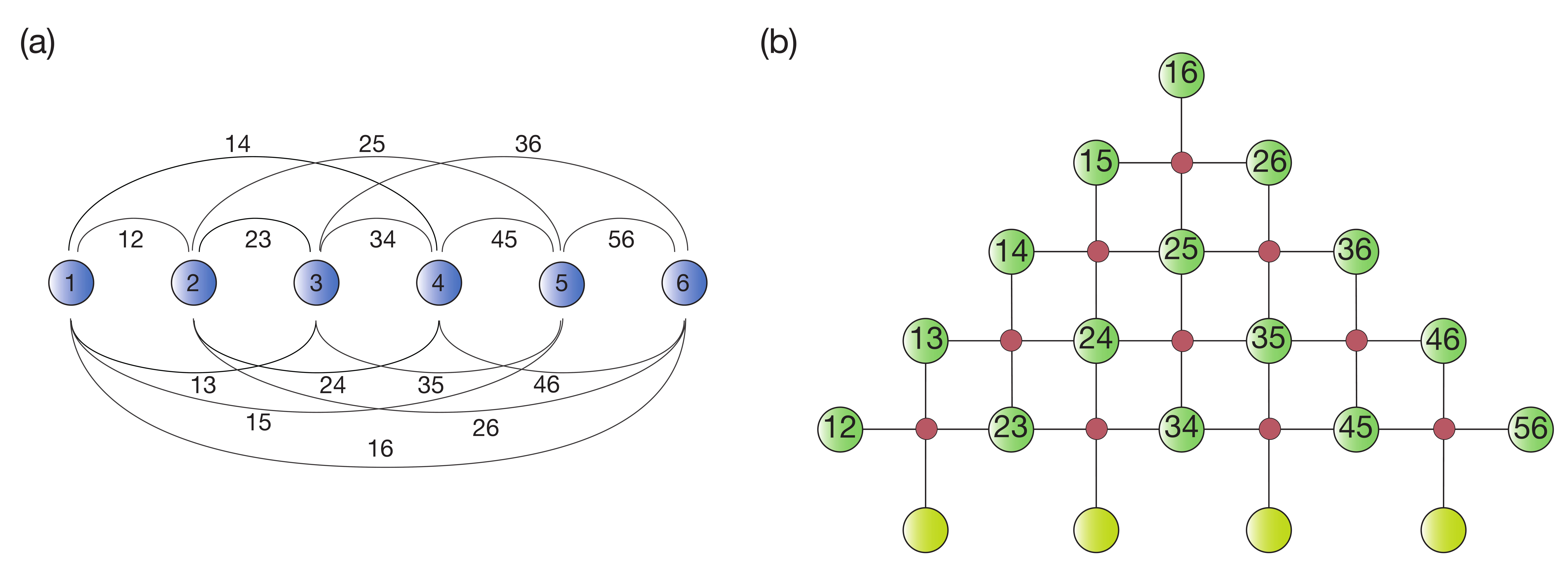}}
\caption{Illustration of the PAQC encoding technique. a) The system of spins for the spin-glass annealing paradigm with all-to-all Ising Interactions. Ising pair interactions are symbolised by lines and labeled by the coupled spins. b) The pyramid shaped, cubic lattice of spins of the PAQC scheme where the parallel or antiparallel alignment of spins $i$ and $j$ of the spin-glass annealing paradigm is encoded in the state of the spin $(i \,j)$. 4-body Ising interactions act on the unit cells of the PAQC spin system symbolised by red circles.}
\label{fig:review}
\end{figure}

In the spin-glass paradigm of AQC one uses a spin-glass with on-site fields and all-to-all Ising interactions (c.f. \fref{fig:review} a),
\begin{equation}
H_{\text{SG}}= \sum\limits_i \left(h_{z,i} \tilde{\sigma}_z^{(i)} + h_{x,i}\tilde{\sigma}_x^{(i)}\right) + \sum\limits_{i,j} J_{i,j} \tilde{\sigma}_{z}^{(i)}\tilde{\sigma}_{z}^{(j)}\,.
\end{equation}
Here, $\tilde{\sigma}_{\{x,y,z\}}^{(i)}$ are the Pauli operators of spins in the spin glass annealing paradigm. The parameter space in spin-glass annealing is given by the set of all on-site fields ($h_{z,i}$ and $h_{x,i}$)  and Ising interaction strengths ($J_{i,j}$). The initial point in parameter space could be the point of dominating on-site transverse fields $|h_{x,i}|\gg |h_{z,j}|$  and $|h_{x,i}|\gg |J_{i,j}|$ $\forall i,j$ and the corresponding groundstate would be the fully separable state  $\left|1\right\rangle=\bigotimes\limits_i (\left|\downarrow\right\rangle -\left|\uparrow\right\rangle)/\sqrt{2}$. The on-site fields and Ising interaction strengths are slowly deformed to prepare the spin-glass in the groundstate of $H^{(\text{final})}$ given in parameter space by , $h_{z,i}\gg |h_{x,j}|$  and $ |J_{i,j}| \gg |h_{x,i}|$ $\forall i,j$. The actual values of $|h_{z,i}|$ and $J_{i,j}$ encode the optimisation problem. The spin glass remains in its groundstate, if the sweep is slow enough, and a subsequent readout of all spin states provides the solution to the given optimisation problem. The main obstacle for the construction of a spin-glass annealer is the implementation of tunable all-to-all Ising interactions.

\subsection{Parity Adiabatic Quantum Computing}

In PAQC, a system with a larger number of spins but only local interactions is used to mimic the spin glass system with all-to-all Ising interactions at the end of the adiabatic sweep. This is possible because there exists a unambiguous mapping of eigenstates of the spin-glass system to the lowest energy states of the PAQC system which are separated by a energy gap from the remaining states of the enlarged Hlibert space of the PAQC system.   

The mapping between states  of the spin-glass system with all-to-all connectivity to the states of the PAQC system is accomplished as follows: The Ising interaction $\sigma_{z,i}\sigma_{z,j}$ has two twofold degenerate eigenvalues: $+1$ and $-1$ corresponding to parallel ($\left|\uparrow\uparrow\right\rangle$ and $\left|\downarrow\downarrow\right\rangle$) and antiparallel ($\left|\uparrow\downarrow\right\rangle$ and $\left|\downarrow\uparrow\right\rangle$) alignment of the spins. In the PAQC system the state of this Ising interaction gets represented by a single spin, say state $\left|\uparrow\right\rangle$ corresponds to a parallel configuration and $\left|\downarrow\right\rangle$ to a antiparallel configuration of spins. This already provides us with an unambiguous mapping from states of the spin-glass with all-to-all connectivity with $N$ spins to the states of the PAQC system with $K=N(N-1)/2$ spins. However the larger Hilbert space of the PAQC system already prohibits an unambiguous mapping of all states to corresponding states of the spin-glass Hamilton operator. The effective reduction of the Hilbert space of the PAQC system is accomplished with constraints. Constraints are terms in the Hamilton operator that afflict large energy penalties on configurations of the PAQC systems that do not have a counterpart in the spin-glass system. 
The PAQC system is a spatial arrangement of the spins which enables only local constraint, i.e. energy penalties involving only neighbouring spins  (c.f. \fref{fig:review} b) and \cite{PROPOSAL}). It consists of a cubic lattice of spins with overall pyramid geometry where the constraints act locally on every unit cell of the cubic lattice. 
The Hamilton operator of PAQC has the form  
\begin{equation}
	H_{\text{PAQC}} = A(t) \sum_i^K \sigma_x^{(i)} + B(t) \sum_i^K J_i \sigma_z^{(i)} + C(t) \sum_l^L \sigma_z^{(l,n)}\sigma_z^{(l,s)}\sigma_z^{(l,e)}\sigma_z^{(l,w)}.
	\label{eq:originalH}
\end{equation} 
Here, $\sigma_{\{x,y,z\}}^{(i)}$ are the Pauli operators spins in the PAQC system. The terms $A(t)$, $B(t)$, and $C(t)$, are three independent schedule functions that define the annealing protocol. $A(t)$, the schedule function of the initial Hamiltonian, is tuned from its maximal value $A_{\text{max}}$ to $0$ during the adiabatic sweep.  $B(t)$, the schedule function of the Hamiltonian that encodes the optimization problem, is tuned from $0$ to its maximal value $B_{\text{max}}$ during the adiabatic sweep. $C(t)$ is the schedule function of the constraint term. $C(t)$ can be switched from $0$ to $C_{\text{max}}$, with $C_{\text{max}}$ dominating every other energy scale in the system. However, PAQC also opens the possibility to keep the constraints ``always-on''. The constraints are implemented with 4-body Ising interactions involving the north- $(l,n)$, east- $(l,e)$, south- $(l,s)$ and west- $(l,w)$ spins of every plaquette $l$. They are constraints because at least at the end of the adiabatic sweep they are the dominating energy contribution and all dynamics is restricted to reside in their groundstate-manifold.

In this way the technological challenge of implementing tunable all-to-all Ising interactions has been changed to implementing local 4-body Ising constraints. In the following we show how to further decompose general many-body Ising constraints to two-body Ising interactions with the help of ancilla spins.

\subsection{Decomposition of Constraints}
The constraint decomposition follows a 2-step process: (i) We present a general recursive technique to decompose many-body Ising constraints to 3-body Ising constraints with ancilla spins and use this technique to decompose the 4-body Ising constraint of PAQC to two 3-body Ising constraint, and (ii) we show a decomposition of a  3-body Ising constraint  with 2-body Ising interactions with one ancilla spin. This constraint decomposition technique follows a similar logic to the embedding described above. However instead of embedding the whole eigensystem of an operator in another, we only need to embed the groundstate-manifold of the constraint and make sure that the groundstate-manifold of the embedding constraint is still separated by a sufficient gap from the rest of the states.   

\paragraph{Decomposing many- to 3-body constraints}
The many-body Ising interaction $C\prod_{i=1}^{M} \sigma_z^{(i)}$, has two multiply degenerate eigenvalues $1$ and $-1$ corresponding to product states of an even or odd number of spins in the spin down state respectively. In the following we call them even- or odd-parity states respectively. Depending on the sign of the constraint strength $C$ either the states with even or parity are the groundstate-manifold, i.e. are ``allowed'' under the constraint. Given a set of qubits in eigenstates of their respective $\sigma_z$ operators we could detect the parity of the state by either detecting the parity of the whole set or subdividing the set into two subsets, $A$ (spins $1$ through to $k$) and $B$ (spins $k+1$ through to N), and detecting the parity of the two subsets. If the two subsets have the same parity, the parity of the set is even and vice versa. We can translate this principle into the domain of Ising interactions with the help of an ancilla spin $(a)$,
\begin{equation}\label{eq:decompgeneral}
C\prod\limits_{i=1}^N\sigma_z^{(i)} \equiv \pm |C| \left(\sigma_z^{(a)}\prod\limits_{i=1}^k\sigma_z^{(i)}-\text{sgn}(C)\sigma_z^{(a)}\prod\limits_{i=k+1}^N\sigma_z^{(i)}\right)
\end{equation}
The equivalence ``$\equiv$'' is here defined as the existence of a one-to-one mapping between the groundstate-manifolds in the sense that the states are identical for the non-ancilla spins. Additionally all remaining states of the embedding constraint have to be separated by a gap from the groundstate-manifold. The overall sign of the embedding constraint above is actually arbitrary which is due to the fact that any gauge transformation that interchanges the meaning of ``up'' and ``down'' states of the ancilla spin leaves the physics invariant however changes the overall sign of the embedding constraint. The constraint decomposition technique is valid for odd $(C>0)$ and even $(C<0)$ parity constraints. The 4-body Ising constraint of PAQC is a even party constraint which is why we have chosen to further illustrate even parity decomposition. The reasoning for odd parity constraints parallels the reasoning given in the following. 
 
 The groundstate manifold of the embedding constraint for an even parity constraint $(C<0)$  is characterised by a odd parity for the subset A including the ancilla spin $(a)$ and the subset B including the same ancilla spin $(a)$. If the ancilla spin $(a)$ is in the spin up state, then both of the subsets A and B have to have odd parity in order for the state to be in the groundstate-manifold. If the ancilla spin $(a)$ is in the spin down state both subsets have to have even parity. The ancilla spin $(a)$ therefore ``communicates'' the parity information between subsets. Therefore the set of spins not including the ancilla spin behaves as if the full many-body Ising constraint were implemented without the need for any perturbative elimination of the ancilla spin. The gap of the embedding constraint above, c.f. equation \eref{eq:decomprule}, is actually of the same size as the gap of the constraints involving the subsets A and B  by virtue of the non-perturbative nature of the decomposition. This opens the door to very strong effective Ising constraints.

\begin{figure}[tb]
\centerline{\includegraphics[width=16cm]{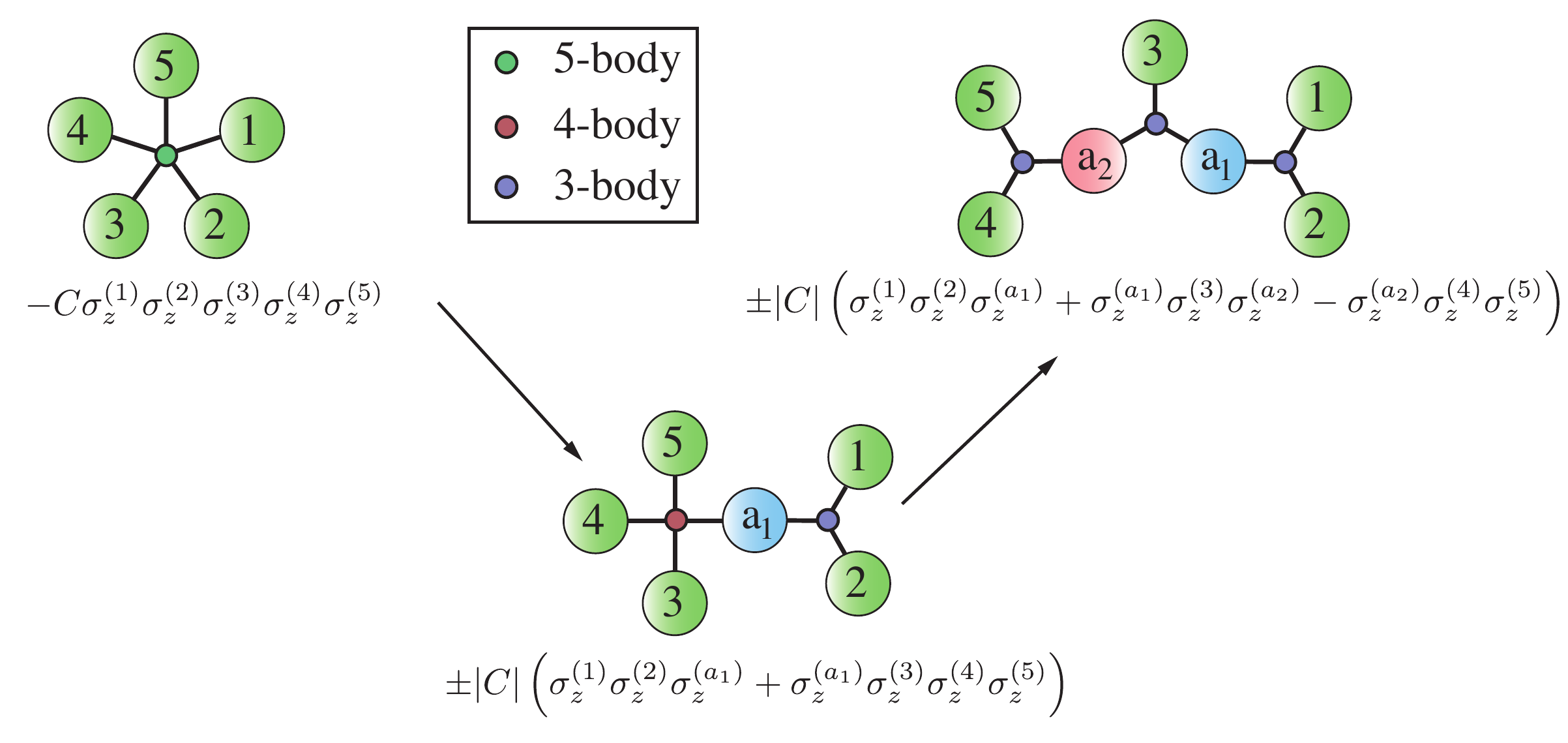}}
\caption{(a) Example of the recursive decomposition of 5-body Ising constraints into a ``tree'' of 3-body Ising constraints. The example starts with a plaquette of 5 qubits that interact via the 5-body Ising even parity constraint $-C \sigma_z^{(1)}\sigma_z^{(2)}\sigma_z^{(3)}\sigma_z^{(4)}\sigma_z^{(5)}$ (top left).  Using Eq. \ref{eq:decompgeneral} we split the system into two parts $\{1,2\}$ (right branch) and $\{3,4,5\}$ (left branch). Adding the ancilla $(a_1)$ results in a 3-body Ising constraint and a 4-body Ising constraint (bottom middle). The three-body Ising constraint can not decomposed any further with this technique and is a ``leaf'' of the tree. However the 4-body Ising constraint is further decomposed into two 3-body Ising constraints with another ancilla spin $(a_2)$ (top right). }
\label{fig:tree}
\end{figure}

To summarize: with the ancilla spin $(a)$ one can decompose a constraint of order $N$, i.e. involving N spins, into two constraints of order $k+1$ and $N-k+1$. This decomposition can be iterated recursively with both subsystems thus one can decompose any constraint of arbitrary order $N$ into a set of coupled three-body constraints. The recursive sequence of decomposition steps generates a tree structure which is a highly desirable feature for two-dimensional setups like the ones used in circuit quantum electrodynamics because qubits are connected without any crossings and no air-bridges are needed.

\Fref{fig:tree} depicts an example of this recursive algorithm for the decomposition of a 5-body constraint. We start by splitting the 5-body term into a right branch with subset $\{1,2\}$ and a left branch with subset $\{3,4,5\}$. In each branch we add the ancilla spin $(a_1)$ resulting in two terms, a three-body constraint in the right branch and a 4-body constraint in the left branch. The left branch is finished as three-body constraints cannot be further decomposed with this scheme. We continue with the left branch and split the 4-body constraint into two subsystems, containing $\{4,5\}$ and $\{a_1,3\}$, respectively. Adding the ancilla qubit $(a_2)$ on both sides results in a three-body constraint in all leafs of the tree which terminates the procedure. After joining the leafs, the final decomposed 5-body constraint is depicted in \Fref{fig:tree}(b). 

\paragraph{Decomposing 3- to 2-body constraints}
The second decomposition step (ii) aims at replacing the remaining 3-body constraints by pair interactions including another ancilla spin. The above recursive decomposition cannot be applied to three-body terms, as one subbranch would again result in a 3-body constraint. However, it is possible to construct a set of interactions that feature the same degenerate ground state as the 3-body Ising constraint if we ignore the state of the ancilla spin for the moment. There are actually many possible combinations of pair interactions that share the same degenerate ground-state. We chose the particular solution, where (i) all interactions have the same sign and (ii) there are no crossings [depicted in Fig. \ref{fig:illustration}(b)]

\begin{eqnarray}
	\label{eq:l3}
	\sigma_z^{(1)}\sigma_z^{(2)}\sigma_z^{(3)} &\equiv & \sigma_z^{(1)}\sigma_z^{(2)} + \sigma_z^{(2)}\sigma_z^{(3)} + \sigma_z^{(3)}\sigma_z^{(1)}\\ \nonumber
	 &+& \sum_{i=1}^3 [2 \sigma_z^{(i)}\sigma_z^{(a)} - \sigma_z^{(i)} ] - 2 \sigma_z^{(a)}.
\end{eqnarray}
Here, the symbol "$\equiv$" quotes the definition of equivalence for constraints given above. Using Eq.~(\ref{eq:l3}) together with Eq.~(\ref{eq:newH}) the parity quantum annealing scheme \cite{PROPOSAL} can be implemented with pair interactions only. The full layout of interactions and ancillas is shown in Fig. \ref{fig:illustration}(c).

\subsection{PAQC with pair Ising interactions}
Let us now apply the above algorithm to decompose all constraints in the PAQC scheme. The first step (i) a trivial tree in this case. We simply split the four body constraint into two three body constraints,
\begin{equation}
 \label{eq:fourtothree}
-C\sigma_z^{(n)}\sigma_z^{(e)}\sigma_z^{(s)}\sigma_z^{(w)} \to 
 - C\left(\sigma_z^{(n)}\sigma_z^{(e)}\sigma_z^{(a)} + \sigma_z^{(a)}\sigma_z^{(s)}\sigma_z^{(w)} \right)
\end{equation}
Plugging Eq.~(\ref{eq:fourtothree}) into Eq.~(\ref{eq:originalH}) results in the adiabatic protocol 
\begin{eqnarray}
	\label{eq:newH}
	H(t) &=& A(t) \sum\limits_i^K \sigma_x^{(i)} + B(t) \sum_i^{K+M} J_i \sigma_z^{(i)} +\\ \nonumber
	 &&C(t) \sum_l^{L} \left( \sigma_z^{(n)}\sigma_z^{(w)}\sigma_z^{(a)} + \sigma_z^{(a)}\sigma_z^{(e)}\sigma_z^{(s)}\right).
\end{eqnarray}
Here, we add one ancilla spin for each plaquette which makes a total of $M=(N-2)^2/2$ spins as shown in Fig. \ref{fig:illustration}. Each of the ancilla spins is shared by two 3-body Ising interactions. Compared to the implementation with 4-body Ising constraints, the number of constraints doubled in the above realization with 3-body Ising constraints. In the second step (ii), each 3-body term in Eq.~(\ref{eq:newH}) is replaced by the right hand side of Eq.~(\ref{eq:l3}) which results in the final layout with pair-interaction only depicted in Fig. \ref{fig:illustration}(c). Note, that this scheme may be optimized by combining pairs of spins [Fig. \ref{fig:illustration}(c) (yellow color)] which results in a layout with less spins but introduces crossings of interactions. 

\subsection{Adiabatic protocol}

The implementation of Eq.~(\ref{eq:originalH}) together with Eq.~(\ref{eq:l3}) contains only pair interactions on a 2D geometry without crossings. In the following, we illustrate the applicability of this implementation for an adiabatic protocol for an individual 4-body Ising constraint as shown in Fig. \ref{fig:illustration}(a). The protocol is as follows: The system is prepared in the trivial ground state of $H_0^a$, which contains the $4$ spins plus $3$ ancilla spins [see Fig. \ref{fig:illustration}(a)]. Then the local field and constraint terms are adiabatically switched on, while the the $\sigma_x$ term is switched adiabatically off. 

\begin{figure}[htb]
\centerline{\includegraphics[width=16cm]{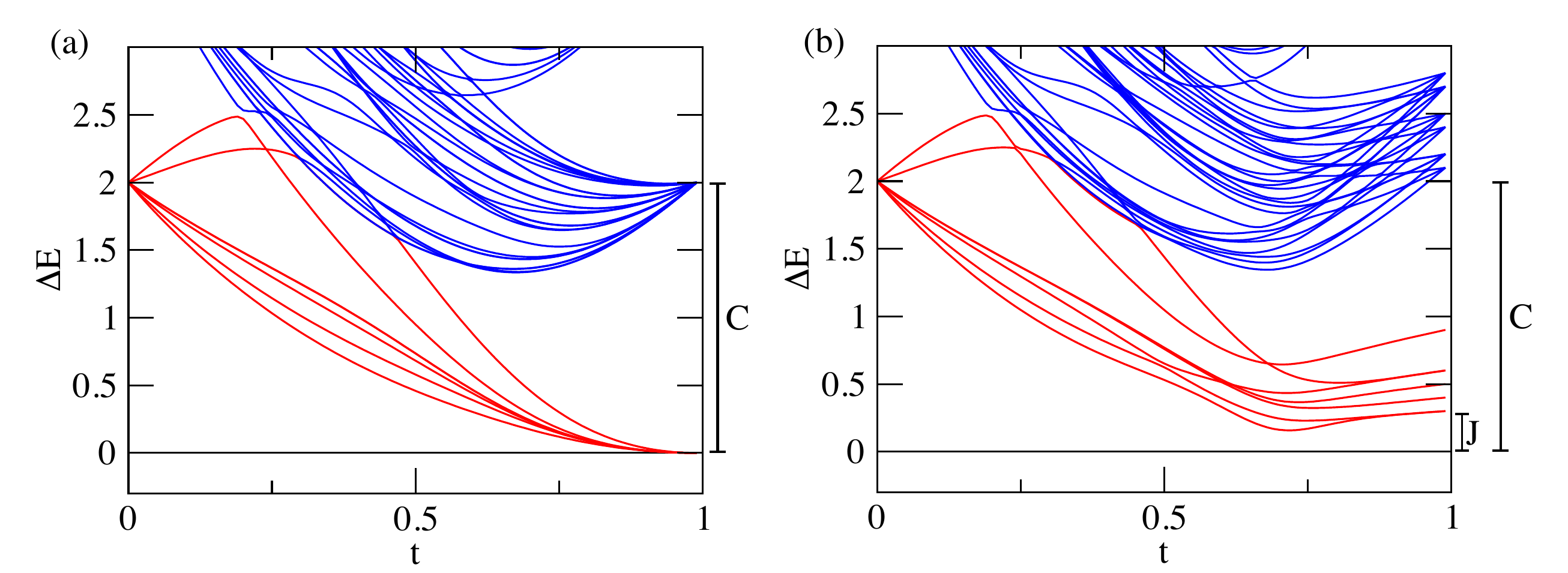}}
\caption{ (a) Time-dependent spectrum of a 4-body term represented by pair interactions [Fig. \ref{fig:illustration}(b)]. In the absence of a programmable local field, all constraint-satisfying states (red) collapse to the degenerate ground state (black). Constraint-violating states (blue) are separated by $C$. The finial state is the superposition of all constraint-satisfying configurations. (b) Spectrum during the sweep with local random fields with strength $|J|=0.2$. The degeneracy of the ground state is lifted and the solution of the optimization problem is the ground state. Note, that the constraint-violating terms are shifted by at least $C$.}
\label{fig:spectrum}
\end{figure}

In the absence of local programmable fields, the result is a superposition of all $8$ constraint-satisfying configurations. The time-dependent spectrum of this ideal sweep is shown in Fig. \ref{fig:spectrum}(a). In the presence of local programable fields, the final ground state is the solution of the optimization problem. The local field terms  $|J|$ lift the degeneracy of the final state  [see. Fig. \ref{fig:spectrum}(b)]. 

\begin{figure}[htb]
\centerline{\includegraphics[width=8cm]{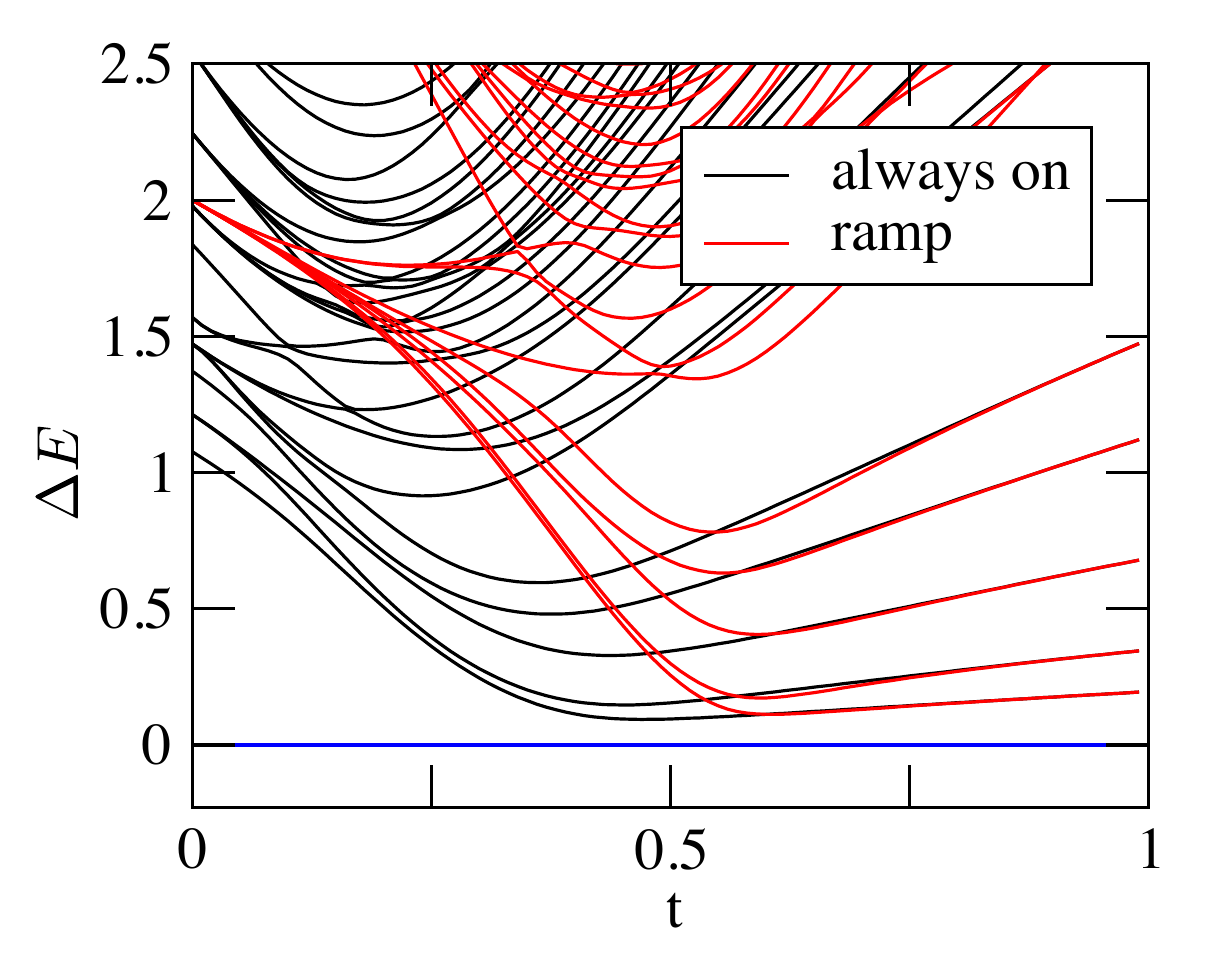}}
\caption{Time-dependent spectrum of the "always on" switching protocol (black) and a ramp of interactions (red) for a typical instance of $J_{ij}$. }
\label{fig:alwayson}
\end{figure}

The time-dependent Hamiltonian Eq.(\ref{eq:originalH}) introduced in Ref. \cite{PROPOSAL} contains 3 independent schedule functions $A(t)$, $B(t)$, and $C(t)$. The only condition for the adiabatic sweep is that the final state with $A(T)=0$ and $B(T) = C(T) = 1$ can be adiabatically reached from the initial state. The main error source in both cases are non-adiabatic Landau-Zener transitions from the ground state to higher states. This opens the possibility for several combinations of sweeps. (i) The protocol switching $A(0) = 1$ to $A(T)=1$, and $B(0) = C(0) = 0$ to $B(T) = C(T) = 1$ is the sweep which mimics to the original formulation of adiabatic quantum optimization. (ii) Because $B$ and $C$ can also be switched independently, the parity scheme also opens the possibility for an "always on" scheme, where $C(t)=1$ for all times. This allows one to implement adiabatic quantum optimization with constant interactions while only local fields are switched. The two protocols are compared in Fig. \ref{fig:alwayson}. Note, that the always-on implementation has large experimental advantages as only local fields are tuned. However, preparation of the initial state in this scenario may be more complicated compared to the ramp protocol.  

\section{Transmon Implementation}

We aim at exploiting the large coherence times of Transmon qubits \cite{WALLRAFF,KOCH,SCHOELLKOPF,MARTINIS} for applications in adiabatic quantum computing. In particular, with the aim to implement the PAQC scheme with Transmons one needs a spin Hamiltonian with (i) individually tunable transveral and  (ii) logitudinal fields and (iii) Ising pair interactions that are large compared to the onsite energies. These terms are not naturally available in Transmons. With a microwave drive we introduce (i)  fully tunable longitudinal fields  and  (ii)  transversal fields in a frame rotating with the microwave drive. (iii) The required Ising pair interaction is implemented based on Josephson ring modulators (JRM) \cite{Bergeal2010a} which improves the attainable coupling strengths with respect to a previous proposal \cite{neumeier,jin}, making the interaction the dominant energy scale. In the following we derive (i)-(iii) in detail.

\subsection{Transmon}
The Transmon has been developed on the basis of charge qubits as a improvement on their coherence times. However, while the charge qubit provides a natural mapping to a spin Hamilton operator with longitudinal and transversal fields, the Transmon is more appropriately described as a harmonic oscillator with small nonlinearity. In the following we show how the longer coherence times of the Transmon are a direct consequence of the lost transversal field term and how to combine the transversal field term and the long coherence times in the following subsection.

A charge qubit is an arrangement of two superconducting islands which are connected by a Josephson junction. The state of the Cooper pair condensate on the two islands is completely  described by a canonical conjugate pair of operators: The number of Cooper pairs $N$ that have tunneled through the Josephson junction starting from the electrical neutral state  and the phase difference of the Cooper pair condensate between the two islands $\phi$. The Hamilton operator of the charge qubit is given by the sum of the electrical energy stored in the capacitor ($C$) that is formed by the two islands and the inductive energy of the Josephson junction,
\begin{equation}
H_{\text{cq}}= 4 E_C (N-n_g)^2 - E_{J}\cos(\phi)
\end{equation}
where $E_C= e^2/2C$ is the charging energy with the elementary charge $e$, $E_J$ the Josephson energy and $n_g= C_g V_g / (2e)$ the Cooper pair equivalent of an external potential $V_g$ with capacitance $C_g$. Here, $n_g$ can be a externally applied classical field or the quantum electrical field of uncontrolled sources as well as quantum fields from coupled resonators or other charge qubits. If we ignore the Josephson energy for the moment, the states of the charge qubit as a function of the externally applied classical electric field $n_g$ are parabolas with origin at integer $n_g$ and degeneracies for half integer $n_g$. Each parabola is the energy of a state $\left|n\right\rangle$ with exactly $n$ Cooper pairs. The degeneracy at half integer $n_g$ is lifted by the introduction of the Josephson energy. This is possible because of the tunneling of Cooper pairs between the islands through the Josephson junction. Around this avoided crossing in the spectrum, one can truncate the Hilbert space of the charge qubit to the two states of exactly defined Cooper pairs and thereby accomplish a mapping to a effective spin Hamilton operator with longitudinal field given by the externally applied classical electrical field $h_z=2E_C(1-2n_g)$ and transversal field given by the Josephson energy $h_x=E_J/2$. The qubit states corresponding to spin ``up'' and ``down'' states are states of well defined Cooper pairs $\left|n\right\rangle\to\left|\uparrow\right\rangle$ and $\left|n+1\right\rangle \to \left|\downarrow\right\rangle$ away from the avoided crossing. Exactly in the middle of the avoided crossing the states are symmetric and anti-symmetric superpositions of states of well defined charge,  $(\left|n\right\rangle+\left|n+1\right\rangle)/\sqrt{2}\to\left|\uparrow\right\rangle$ and $(\left|n\right\rangle-\left|n+1\right\rangle)/\sqrt{2}\to\left|\downarrow\right\rangle$. Because the charge is not well defined at the avoided crossing, the charge qubit shows increased resilience to external fluctuations in the electric field. The Transmon improves on this idea by generating a universal avoided crossing for a very strong Josephson tunneling which mixes all charge states $\left|n\right\rangle$, $E_J/E_C\gg1$. In this parameter regime the difference between the maximal and minimal energy as a function of $n_g$ decreases exponentially as a function of $E_J/E_C$ while the nonlinearity, i.e. the difference between the energy groundstate  to the first exited state $E_1-E_0$ and the first- to the second excited $E_2-E_1$ state decreases polynomially. Here $E_n$ are the eigenenergies of the Transmon. The Transmon thereby combines the advantages of a universal sweet spot with sufficient nonlinearity, necessary for individual addressing of its eigenstates with microwave drives. 

The eigenstates of the Transmon are characterized by small zero-point fluctuations in the phase by virtue of $E_J\gg E_C$ which is why it is appropriate to truncate the Josephson energy which is given by the cosine of the phase to fourth order,
\begin{eqnarray}\label{eq:chargeQB}
H_{\text{cq}}&=& 4 E_C (N-n_g)^2 - E_{J}\cos(\phi)\\
& \approx& 4 E_C (N-n_g)^2 - E_{J} + \frac{\phi^2}{2} - E_J \frac{\phi^4}{24}
\end{eqnarray}
The first two terms of the truncated Hamilton operator can be identified by the Hamilton operator of a harmonic oscillator. The externally applied $n_g$ represents a shift in space for the harmonic oscillator that does not change the eigenenergies which shows the existence of the universal ``sweet spot'' \cite{KOCH}. We may therefore as well set $n_g\to 0$ in the following and introduce lowering and raising operators to describe the number of Cooper pairs and phase of the Transmon,
\begin{eqnarray}
N&=\frac{i}{2}\left(\frac{E_{J}}{2E_C}\right)^{\frac{1}{4}}(a-a^{\dag}) \\
\phi&=\left(\frac{2E_C}{E_{J}}\right)^{\frac{1}{4}}(a+a^{\dag})\,.
\end{eqnarray} 
The truncated Hamilton operator of the Transmon can be approximated with the help of an rotating wave approximation that transforms the nonlinear fourth order phase term to a Kerr nonlinearity,
\begin{equation}
4 E_C (N-n_g)^2 - E_{J} + \frac{\phi^2}{2} - E_J \frac{\phi^4}{24} \approx \sqrt{8 E_J E_C} a^\dag a - \frac{E_C}{2} a^\dag a^\dag aa = H_{1T}\,.
\end{equation}
Here we shifted the groundstate energy to zero and neglected a small renormalization of the Transmon energy coming from the normal ordering procedure of the fourth order phase term.  The Transmon is therefore more appropriately described as a nonlinear harmonic oscillator and as a consequence of the universal ``sweet spot'' there is no mapping to a spin Hamilton operator with a transversal field. In the following subsection we show how to reintroduce a transversal field in a rotating frame by a constant microwave drive.

\subsection{Rotating Frame}
A viable way to reintroduce a transversal field is to excite the Transmon with a constant microwave drive of frequency $\omega_d$, strength $A$, and Hamilton operator $H_{\text{drive}}=A (a e^{i\omega_d t}+a^{\dag} e^{-i\omega_d t})$. In a frame rotating with this microwave drive, $U=\exp(-i \omega_d t a^{\dag}a)$ the transverse field term is reintroduced,
\begin{equation}
U^{\dag}(H_{\text{1T}}+H_{\text{drive}}) U - i\hbar U^{\dag}\dot{U}=\\
2J a^{\dag}a -\frac{E_C}{2}a^{\dag}a^{\dag}aa+ A (a+a^{\dag})\,,
\end{equation}
where $2J=\sqrt{8E_CE_J}-\hbar\omega_d$ is the energy equivalent of the frequency difference between Transmon and microwave drive. Here, $\hbar$ is the reduced Planck constant. The symbol $J$ is chosen in anticipation for the use in the PAQC scheme as the onsite longitudinal field that encodes the coupling strength of the associated spin-glass annealer. The quantum annealing processor is operated in a regime where we can neglect occupation of states higher than the first excited state, which is ensured by choosing a driving strength smaller than the nonlinearity $A<E_C$. Therefore, we may project the Hamilton operator to the qubit subspace to get the effective Hamilton operator in the qubit subspace,
\begin{equation}
H_{\text{1T,Qbit}}=J\sigma_z+ A\sigma_x\,.
\end{equation}
To summarize: in a frame rotating with the microwave drive the effective longitudinal field is defined by the energy equivalent of the frequency difference between the Transmon and the microwave drive $J$ which allows longitudinal fields smaller in magnitude than the Ising pair interaction presented below. Additionally a transversal field, given by the strength of the microwave drive $A$, is reintroduced. 

\subsection{Ising pair interaction}
Transmon qubits are naturally coupled via their electric field \cite{WALLRAFF}. At first we shortly illustrate why this capacitive coupling is linear in the Transmon field operators while the desired Ising pair interaction is of fourth order. Afterwards we present  how to implement the Ising pair interaction with the nonlinearity provided by Josephson junctions.

For capacitively coupled Transmons, $n_g$ as given above in equation \ref{eq:chargeQB} of one Transmon $(a)$ is a function of the electric field of another Transmon $(b)$ and their mutual capacitance $C_c$. The interaction term in the Hamilton operator is, 
\begin{equation}
H_{\text{int,cap}}  = \frac{(2e (N_1-N_2))^2}{2C_c} \approx g_{\text{cap}} (a^\dag b + a b^\dag)\,.
\end{equation}
Where we assumed the validity of the Transmon approximation and performed a rotating wave approximation valid for small coupling strengths $g< \omega_a+ \omega_b$. Here, $\omega_a$ and $\omega_b$ are the frequencies of the two Transmons while $a$ and $b$ are their field operators. The natural capacitive coupling of two Transmons therefore provides us with a linear exchange interaction.
However, the fundamental building block of the 4-body Ising constraint which is needed for PAQC is, by virtue of the  decomposition techniques described above (c.f. Figure \ref{fig:illustration}(c)), the Ising pair interaction $\sigma_z^{(a)}\sigma_z^{(b)}$. 
In terms of field operators of the Transmon, $a$ and $a^{\dag}$, every $\sigma_z$ operator is already a quadratic operator because $2 (a^{\dag}a-1/2)$ corresponds to $\sigma_z$ in the qubit subspace and consequentially  our desired Ising pair interaction term is of fourth order. 
Therefore this interaction has to be implemented with the only nonlinear element at our disposal for superconducting qubits, the Josephson junction. One possibility is to connect an island of the first Transmon to one of the islands of the other, which results in a coupling energy given by,
\begin{equation}
H_{\text{int,ind}}  = E_{J_c} \cos(\phi_1-\phi_2)\approx g_{\text{ind}} (a^\dag b + a b^\dag) + g_{\text{nonlin,ind}}(a + a^\dag-b-b^\dag)^4\,.
\end{equation}
Here we again assumed the validity of the Transmon approximation and performed a rotating wave approximation to get the linear exchange interaction. Additionally we truncated the cosine of the Josephson junction at fourth order which is valid given the Transmon approximation. In addition to our desired fourth order nonlinear coupling we get a linear exchange interaction.
It is possible to cancel the linear terms provided by a Josephson junction with a parallel capacitive interaction \cite{neumeier,jin}  because $g_{\text{cap}}$ and $g_{\text{ind}}$ have a different sign. However, this scheme limits the interaction energy, because the coupling capacitance has to be considerably smaller than the Transmon's capacitance to inhibit unwanted long-range interactions.

We present another possibility to couple Transmons with Josephson junctions in the form of a JRM \cite{Bergeal2010a} where the linear coupling vanishes for symmetry reasons. The interaction energies exceed the coupling energies attainable by the capacitively shunted Josephson junction scheme by at least an order of magnitude without introducing unwanted long-range interactions. 

Four superconducting islands joined to a ring by identical Josephson junctions define a JRM (c.f. Appendix A). Its energy is proportional to the product of the cosines of three orthogonal modes, if the flux threaded through the JRM loop vanishes. The JRM's modes are comprised of two differential modes involving opposing superconducting islands $\varphi_x=\varphi_1-\varphi_3$, $\varphi_y=\varphi_2-\varphi_4$ and a third mode involving all islands $\varphi_z=\varphi_1-\varphi_2+\varphi_3-\varphi_4$. Here, $\varphi_i =\int_{-\infty}^t V_i dt$ is the flux variable defined as the time integral of the electrical potential $V_i$ of island $i$. It is connected to the phase variable introduced above by $\varphi_i = \phi \varphi_0$ with the reduced magnetic flux quantum $\varphi_0=\hbar/(2e)$.

We associate the two differential modes $\varphi_x$ and $\varphi_y$ of the JRM with the modes that register the two qubits $\varphi_a$ and $\varphi_b$ by connecting them with conducting leads. For the sake of simplicity we assume all JRM junctions to be equal with Josephson energy $E_\jrm$, although our setup tolerates the usual fabrication inaccuracies (c.f. Appendix B). The qubit capacitances $C_J$ are also assumed to be equal although unequal qubit capacitances do not alter the main result. The qubit Josephson junctions can be implemented as direct current superconducting quantum interference devices (dc-SQUID) and are modelled as tunable Josephson energies $E_{Ja}$ and $E_{Jb}$. Note that the Transmons might as well be implemented with fixed frequencies provided we can individually change the drive frequencies. They are the ``nobs'' of the quantum annealing processor that need to be adjusted in order to encode the problem we want to solve. 
\begin{figure}[tb]
\centerline{\includegraphics[width=10cm]{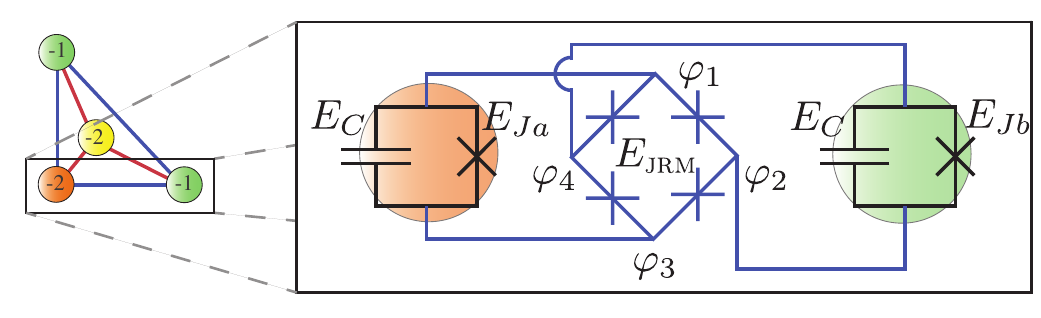}}
\caption{Illustration of the circuit diagram for the two-qubit building blocks of the quantum annealing processor. As all interactions are the same, this is the only required building block. Two Transmons with charging energy $E_C$ and Josephson energies $E_{Ja}$ and $E_{Jb}$ respectively are coupled by a Josephson ring modulator with Josephson energy $E_\jrm$}
\label{fig:JRMcircuit}
\end{figure}
Contrary to the typical use case of the JRM with $\varphi_{ext}\neq 0$ we are interested in the case $\varphi_{ext}=0$. The  two-Transmon Lagrangian with the corresponding JRM contribution reads (for a full derivation of the Lagrangian of the JRM c.f. \ref{app:symJRM}),
\begin{eqnarray}
L'_{2T}&=&\frac{C_J}{2}\dot{\varphi}_a^2 + E_{Ja}\cos(\frac{\varphi_a}{\varphi_0}) + \frac{C_J}{2}\dot{\varphi}_b^2 + E_{Jb}\cos(\frac{\varphi_b}{\varphi_0}) + \\
&& + 4E_\jrm\cos(\frac{\varphi_a}{2\varphi_0})\cos(\frac{\varphi_b}{2\varphi_0})\cos(\frac{\varphi_z}{2\varphi_0})\,.
\end{eqnarray}
Notice that the third mode $\varphi_z$ of the JRM does not possess any capacitive term. In a mechanical picture this mode is a massless particle moving in a one-dimensional potential that depends on the position of other massive particles. Therefore it will immediately adjust itself to the potential minimum. In analogy to the elimination of the coupling degree of freedom for the g-mon \cite{geller} we find,
\begin{eqnarray}
 \frac{d}{dt}\frac{\partial L'_{2T}}{\partial\dot{\varphi}_z}=\frac{\partial L'_{2T}}{\partial\varphi_z}&=&0\\
 \cos(\frac{\varphi_a}{2\varphi_0})\cos(\frac{\varphi_b}{2\varphi_0})\sin(\frac{\varphi_z}{2\varphi_0})&=&0\,,
 \end{eqnarray}
which is fulfilled irrespective of the values for $\varphi_a$ and $\varphi_b$ if $\varphi_z=0$ for all times.   
We Legendre transform the resulting Lagrangian and quantize the theory to get the Hamilton operator,
\begin{eqnarray}
H'_{2T}&=&4 E_C N_a^2 - E_{Ja}\cos(\phi_a) + 4E_C N_b^2 - E_{Jb}\cos(\phi_b) -\\
&&- 4E_\jrm\cos(\frac{\phi_a}{2})\cos(\frac{\phi_b}{2})\,.
\end{eqnarray} 
In preparation for the Transmon approximation we introduce bosonic lowering and raising operators with the following dependency on the Cooper pair and phase operators,
\begin{eqnarray}
N_x&=\frac{i}{2}\left(\frac{E_{Jx}+E_\jrm}{2E_C}\right)^{\frac{1}{4}}(x-x^{\dag}) \\
\phi_x&=\left(\frac{2E_C}{E_{Jx}+E_\jrm}\right)^{\frac{1}{4}}(x+x^{\dag})\,,
\end{eqnarray}
where $x \in \{a,b\}$.
Truncating the Hamilton operator to quartic order, results in the canonical Transmon approximation of a harmonic oscillator mode with a quartic nonlinearity given by the charging energy of the Transmon,
\begin{equation}
H'_{2T}\approx H_{2T}=E_a a^{\dag}a-\frac{E_C}{2}a^{\dag}a^{\dag}aa+
E_b b^{\dag}b-\frac{E_C}{2}b^{\dag}b^{\dag}bb -g(a+a^{\dag})^2(b+b^{\dag})^2
\end{equation} 
with 
\begin{eqnarray}
E_x = \sqrt{8E_C(E_{Jx}+E_\jrm)} \label{eq:tEnergy} \\
g = \frac{E_C}{2}\frac{E_\jrm}{\sqrt{E_{Ja}+E_\jrm}\sqrt{E_{Jb}+E_\jrm}}\label{eq:cStrength}\,.
\end{eqnarray}
As a last step we add the individual microwave drives, introduced above, for both Transmons and transform into a frame rotating with the microwave drives with $U_x=\exp(-i \omega_{d,x} t x^{\dag}x)$ for $x\in\{a,b\}$. Note, that the $\sigma_z$ coupling in Eq.~(\ref{eq:cStrength}) is of the order of the non-linearity of the Transmon. The validity of our projection to the individual qubit subspaces  still only requires the driving strengths $A_a$ and $A_b$ to be small compared to the onsite non-linearity $E_C$. The effective Hamilton operator in the rotating frame projected to the qubit subspace is therefore,
\begin{eqnarray}\label{eq:transmoninteraction}
H_{\text{2T,Qbit}}=A_a\sigma_x^{(a)} + A_b\sigma_x^{(b)} + J_a\sigma_{z}^{(a)} + J_b\sigma_{z}^{(b)} - g\sigma_{z}^{(a)}\sigma_{z}^{(b)}\,,
\end{eqnarray}
where $J_x=E_x-2g-\hbar\omega_{d,x}$  for $x\in\{a,b\}$.

With the longitudinal and transversal field terms and the Ising pair interaction we have all the ingredients to build a Transmon quantum annealer with the PAQC scheme. Errors on the local field terms $\sigma_z$ and $\sigma_x$  are highly reduced in the rotating frame due to $ns$-precision in microwave drives of Transmons \cite{Kelly}. As the optimization problem is encoded in the z-direction of the Hamiltonian, any error in the $\sigma_z$ terms must be smaller than the range of programmability $C \gg |J| \gg \delta_z,k_B T$. Errors arise from the second order processes neglected in the derivation of Eq.~(\ref{eq:transmoninteraction}) that result in additional $\sigma_x\sigma_x$ couplings. Let us also note additional opportunities for implementations with our scheme. The presented approach allows for an implementation of a Transmon quantum annealer with fixed frequency qubits \cite{IBM} which show increased resilience with respect to environmental noise. By driving each Transmon individually $\omega_{d,a}\neq\omega_{d,b}$ and upon changing into the rotating frame one gets an effective time-independent Hamilton operator. One might even conceive of a Transmon quantum annealer where each Transmon is replaced by a capacitor and the JRMs alone provide the inductive energy for the onsite Transmons, as the effective Josephson energy of both Transmons is the sum of the Josephson energy of the Transmon junction and the JRM junction Eq.~(\ref{eq:tEnergy}).

\section{Conclusion}

With the strong resilience to noise, the Transmon qubit opens the possibility to go beyond temperature driven annealing protocols and to study the influence of coherent vs. incoherent processes in an adiabatic sweep. On the implementation level, we showed how to introduce tunable $\sigma_x$ and $\sigma_z$ terms in Transmons as well as $\sigma_z\sigma_z$-interactions from Josephson ring modulators that can be larger than the local field terms. On the encoding level, we developed a recursive constraint decomposition method that allows one to break any combination of classical k-local constraints into coupled 3-body constraints with ancillas. Due to the recursive nature of the algorithm the resulting graph is a binary tree and therefore does not feature any crossings or junctions. The number of ancillas scales linear with $k-3$ and the recursion terminates at 3-body terms. To further decompose 3-body terms to pair interactions we followed a different strategy where only the ground state is degenerate. The result is a system where (i) constraint-satisfying terms are degenerate, (ii) the constraint-violating terms are higher in energy, (iii) the layout has no crossings and (iv) all interactions have the same sign. 

The decomposition in combination with the parity adiabatic quantum computing scheme \cite{PROPOSAL} and the rotating frame Transmon qubit aims at implementing a quantum annealer with full all-to-all programmability from physical two-body interactions and local fields alone. The overhead of the encoding is 3 ancilla-qubits per plaquette. This Transmon quantum annealer provides an alternative to the flux qubit annealer with the prospect of large coherence times that could improve the annealing success probability considerable. Quantum annealing in the rotating frame adds additional flexibility  which allows for studies of the influence of coherence in quantum annealing. However, the proposed scheme also poses questions on how incoherent processes influence the success probability which we plan to address in future work.
\newline

{\it During the course of writing up this paper we became aware of related work: In reference \cite{YingLi} the authors present a PAQC encoding which introduces odd instead of even parity constraints that are easier to implement with pair Ising interactions and a single ancilla spin per constraint. This reduces the number of ancilla spins, however the connectivity graph is not flat. In reference \cite{Warburton} a PAQC implementation based on flux qubits is introduced with 4 ancilla qubits per plaquette.}

\ack We acknowledge fruitful discussions with Gerhard Kirchmair. ML acknowledges support from the Lise-Meitner project M 1972-N27. WL was supported by the Austrian Science Fund (FWF): P 25454-N27 and PZ was supported by ERC Synergy Grant UQUAM and SFB FoQuS (FWF Project No. F4016-N23).

\newpage

\appendix

\section*{Josephson Ring Modulator}
\label{app:symJRM}
The Josephson ring modulator (JRM) as applied in the main text as a coupling device is originally employed in the regime $\varphi_{\circ}=\pi\varphi_0/4$ as a three wave mixing device for readout purposes  \cite{Bergeal2010a}. Here, we provide a derivation for the main JRM characteristics and evaluate the performance of the JRM with unequal Josephson junctions. The JRM is a square composed of large area Josephson junctions $E_{JRM}\gg E_C$ c.f. Fig. \ref{fig:JRMcircuit}. The Lagrangian reads,
\begin{eqnarray}
L_{JRM}&=&E_{JRM}\Big(\cos(\frac{\tilde{\varphi}_1-\tilde{\varphi}_2}{\varphi_0})+\cos(\frac{\tilde{\varphi}_2-\tilde{\varphi}_3}{\varphi_0})+\cos(\frac{\tilde{\varphi}_3-\tilde{\varphi}_4}{\varphi_0})+\\
&&+\cos(\frac{\tilde{\varphi}_4-\tilde{\varphi}_1-\varphi_{\circ}}{\varphi_0})\Big)\,,
\end{eqnarray}
with $\varphi_0=\phi_0/(2\pi)=\hbar/(2e)$ the reduced quantum of flux and the flux enclosed by the ring $\varphi_{\circ}$. Because of the enclosed flux $\varphi_{\circ}$ a ring current is present even in a static- or ground-state $\dot{\tilde{\varphi}}_i=0$ $\forall i$. Therefore all node fluxes $\tilde{\varphi}_i$ may be expressed as the sum of a static $\varphi_i^\dc(\varphi_{\circ})$ and dynamic $\varphi_i^\ac$ part $\tilde{\varphi_i} = \varphi_i^\dc+\varphi_i^\ac$. While the static fluxes are parameters of the circuit, the dynamic node fluxes are the degrees of freedom of the JRM. In order to reveal the full role of the enclosed flux in the Lagrangian it is necessary to determine the static node fluxes $\varphi_i^{\dc}$ as a function of the enclosed flux $\varphi_{\circ}$.
The correct values for the static fluxes can be determined with the help of the Euler Lagrange equations,
\begin{equation}
\underbrace{\frac{d}{dt}\frac{\partial L_{\jrm}}{\partial \dot{\tilde{\varphi}}_i}}_{=0}-\frac{\partial L_{\jrm}}{\partial \tilde{\varphi}_i}|_{\{\tilde{\varphi}_i=\varphi^\dc_i\}} = 0 \quad \forall i
\end{equation}
These equations are the Kirchhoff node rules involving only the inductive branches since no constant current can flow through a capacitor. The set of equations reads,
\begin{eqnarray}
\sin(\frac{\varphi_1^{\dc}-\varphi_2^{\dc}}{\varphi_0})&=&\sin(\frac{\varphi_4^{\dc}-\varphi_1^{\dc}-\varphi_{\circ}}{\varphi_0}) \\
 \sin(\frac{\varphi_2^{\dc}-\varphi_3^{\dc}}{\varphi_0})&=&\sin(\frac{\varphi_1^{\dc}-\varphi_2^{\dc}}{\varphi_0})\\
\sin(\frac{\varphi_3^{\dc}-\varphi_4^{\dc}}{\varphi_0})&=&\sin(\frac{\varphi_2^{\dc}-\varphi_3^{\dc}}{\varphi_0})\\
 \sin(\frac{\varphi_4^{\dc}-\varphi_1^{\dc}-\varphi_{\circ}}{\varphi_0})&=&\sin(\frac{\varphi_3^{\dc}-\varphi_4^{\dc}}{\varphi_0})\,.
\end{eqnarray}
This set of equations suggests a solution with equal static flux drops at all Josephson junctions of the JRM  $\varphi_1^{DC}-\varphi_2^{DC}=\varphi_2^{DC}-\varphi_3^{DC}=\varphi_3^{DC}-\varphi_4^{DC}=\varphi_4^{DC}-\varphi_1^{DC}-\varphi_{\circ}$. Additionally all flux drops summed up around the loop should equal $\varphi_{\circ}$, which implies $\varphi_2^{DC}=(1/4)\varphi_{\circ}$, $\varphi_3^{DC}=(2/4)\varphi_{\circ}$ and $\varphi_4^{DC}=(3/4)\varphi_{\circ}$. This corresponds to a situation of a static ring current of equal strength $I= (E_{JRM}/\varphi_0)\sin(\varphi_{\circ}/(4\varphi_0))$ at each Josephson junction of the JRM. 
The Lagrangian expressed in terms of dynamic node fluxes $\varphi_i^\ac$  is ,
\begin{equation}
L_{\jrm}=E_{\jrm}\left(\text{Co}\cos(\frac{\varphi_{\circ}}{4\varphi_0})-\text{Si}\sin(\frac{\varphi_{\circ}}{4\varphi_0})\right)\,,
\end{equation}
with 
\begin{eqnarray}
\text{Co}&=&\cos(\frac{\varphi_1-\varphi_2}{\varphi_0})+\cos(\frac{\varphi_2-\varphi_3}{\varphi_0})+\cos(\frac{\varphi_3-\varphi_4}{\varphi_0})+\cos(\frac{\varphi_4-\varphi_1}{\varphi_0})\\
\text{Si}&=&\sin(\frac{\varphi_1-\varphi_2}{\varphi_0})+\sin(\frac{\varphi_2-\varphi_3}{\varphi_0})+\sin(\frac{\varphi_3-\varphi_4}{\varphi_0})+\sin(\frac{\varphi_4-\varphi_1}{\varphi_0})
\end{eqnarray}
here and in the following we drop the superscript $\ac$ as there is no danger of confusion. In the main text we apply the same labeling convention. With this shift to the minimum in the inductive potential we restored the circular symmetry of the JRM and employ this symmetry to reformulate the Lagrangian to,
\begin{eqnarray}
L_{\jrm}&=&4E_{\jrm}\big(\cos(\frac{\varphi_x}{2\varphi_0})\cos(\frac{\varphi_y}{2\varphi_0})\cos(\frac{\varphi_z}{2\varphi_0})\cos(\frac{\varphi_{\circ}}{4\varphi_0})-\\
&&-\sin(\frac{\varphi_x}{2\varphi_0})\sin(\frac{\varphi_y}{2\varphi_0})\sin(\frac{\varphi_z}{2\varphi_0})\sin(\frac{\varphi_{\circ}}{4\varphi_0})\big)\,,
\end{eqnarray} 
where $\varphi_x = \varphi_1-\varphi_3$, $\varphi_y=\varphi_2-\varphi_4$ and $\varphi_z=\varphi_1-\varphi_2+\varphi_3-\varphi_4$. In the main text we associate modes $\varphi_x$ and $\varphi_y$ with Transmon modes to couple them via the JRM. Note the reduction in the degrees of freedom for the JRM $\{\varphi_1,\varphi_2,\varphi_3,\varphi_4\}\to\{\varphi_x,\varphi_y,\varphi_z\}$ which stems from the fact that a fourth mode $\varphi_m=\varphi_1+\varphi_2+\varphi_3+\varphi_4$ does not couple to the inductive potential given by the Josephon junctions and may therefore be ignored.
The flux enclosed by the JRM loop $\varphi_{\circ}$ is defined by the externally applied flux  up to multiples of the superconducting quantum of flux ($\varphi_{\circ}= (2\pi \varphi_0) n + \varphi_{ext}$ for $n\in \mathbb{Z}$). This implies that the JRM is likely to minimize its energy for various externally applied fields $\varphi_{ext}$ by trapping or releasing flux quanta. In total there are four distinct states of the JRM corresponding to $n = \left\{0,1,2,3\right\} + 4 n$ for $n\in\mathbb{Z}$. Stable operation of the JRM is only possible in the state that minimizes the total energy. This restricts the tunability of the JRM. However we are only interested in a regime where $\varphi_{ext}=0$ and therefore get stable operation for $\varphi_{ext}=\varphi_{\circ}=0$. 

\section*{Non-symmetric Josephson ring modulator}

The Josephson junctions can not be made identical in a large setup like the quantum annealing chip we propose. Imperfections in the Josephson junctions defining the superconducting interference devices (SQUID) of the Transmons do not have any effects on the workings of the chip since the Transmon frequency is defined by the flux threaded through the SQUID and can be adjusted during the operation of the quantum annealer. Therefore we here investigate implications of imperfect Josephson junctions of the JRM. Lets assume four different Josephson energies $E_{J1}$, $E_{J2}$, $E_{J3}$ and $E_{J4}$ with mean value $E_{JRM}$ and variance $\Delta E_i$, then the Lagrangian reads,
\begin{eqnarray}
L_{\jrm}&=&E_{J1}\cos(\frac{\varphi_1-\varphi_2 + \delta_{1,2}}{\varphi_0})+E_{J2}\cos(\frac{\varphi_2-\varphi_3+\delta_{2,3}}{\varphi_0})+\\
&&E_{J3}\cos(\frac{\varphi_3-\varphi_4+\delta_{3,4}}{\varphi_0})+E_{J4}\cos(\frac{\varphi_4-\varphi_1+\delta_{4,1}}{\varphi_0})\,.
\end{eqnarray}
with the static flux jumps $\delta_{1,2}=\varphi_1^\dc-\varphi_2^\dc$, $\delta_{2,3}=\varphi_2^\dc-\varphi_3^\dc$, $\delta_{3,4}=\varphi_3^\dc-\varphi_4^\dc$ and $\delta_{4,1}=\varphi_4^\dc-\varphi_1^\dc+\varphi_\circ$. The static flux jumps can be determined with the help of the steady state Kirchhoff rules of the JRM,
\begin{eqnarray}
E_{J1} \sin(\frac{\delta_{1,2}}{\varphi_0}) &=& E_{J4} \sin(\frac{\delta_{4,1}}{\varphi_0}) \\
 E_{J2} \sin(\frac{\delta_{2,3}}{\varphi_0}) &= &E_{J1} \sin(\frac{\delta_{1,2}}{\varphi_0})\\
E_{J3} \sin(\frac{\delta_{3,4}}{\varphi_0}) &=& E_{J2} \sin(\frac{\delta_{2,3}}{\varphi_0}) \\
 E_{J4} \sin(\frac{\delta_{4,1}}{\varphi_0}) &=& E_{J3} \sin(\frac{\delta_{3,4}}{\varphi_0})
\end{eqnarray}
For the desired operation mode of very small flux enclosed by the JRM loop $\varphi_{\circ}/\varphi_0\ll 1$ we may linearize the above equations and find the solution $\delta_{1,2}=(1/E_{J1})(\varphi_{\circ}/\alpha)$, $\delta_{2,3}=(1/E_{J2})(\varphi_{\circ}/\alpha)$, $\delta_{3,4}=(1/E_{J3})(\varphi_{\circ}/\alpha)$ and $\delta_{4,1}=(1/E_{J4})(\varphi_{\circ}/\alpha)$ with $\alpha = (E_{J1}^{-1}+E_{J2}^{-1}+E_{J3}^{-1}+E_{J4}^{-1})$. The Lagrangian for the JRM up to second order in the flux enclosed by the loop $\varphi_\circ/\varphi_0$ is expressed in terms of the collective modes $\varphi_x$,$\varphi_y$ and $\varphi_z$,
\begin{eqnarray*}
L_{JRM}=&(E'_{J1}+E'_{J2}+E'_{J3}+E'_{J4})\cos(\frac{\varphi_x}{2\varphi_0})\cos(\frac{\varphi_y}{2\varphi_0})\cos(\frac{\varphi_z}{2\varphi_0})-\\
&-(-\Delta E'_1+\Delta E'_2-\Delta E'_3+\Delta E'_4)\sin(\frac{\varphi_x}{2\varphi_0})\sin(\frac{\varphi_y}{2\varphi_0})\cos(\frac{\varphi_z}{2\varphi_0})-\\
&-(\Delta E'_1-\Delta E'_2-\Delta E'_3+\Delta E'_4)\sin(\frac{\varphi_x}{2\varphi_0})\cos(\frac{\varphi_y}{2\varphi_0})\sin(\frac{\varphi_z}{2\varphi_0})-\\
&-(-\Delta E'_1-\Delta E'_2+\Delta E'_3+\Delta E'_4)\cos(\frac{\varphi_x}{2\varphi_0})\sin(\frac{\varphi_y}{2\varphi_0})\sin(\frac{\varphi_z}{2\varphi_0})\\
&-\frac{\varphi_\circ}{\alpha\varphi_0} 4\sin(\frac{\varphi_x}{2\varphi_0})\sin(\frac{\varphi_y}{2\varphi_0})\sin(\frac{\varphi_z}{2\varphi_0})\,.
\end{eqnarray*} 
Here the energies are renomalized slightly due to the non-zero enclosed flux, $E'_{Ji}=E_{Ji}(1-(1/2)(\varphi_\circ/(E_{Ji}\varphi_0\alpha))^2)$ and $\Delta E'_i=\Delta E_i(1-(1/2)(\varphi_\circ/(E_{Ji}\varphi_0\alpha))^2)$
In the main text we associate modes $\varphi_x$ and $\varphi_y$ with the Transmons with qubit capacitance $C$. The mode $\varphi_z$ is not capacitively shunted and therefore a false degree of freedom and gets eliminated before quantization $\varphi_z\to 0$. The last three terms in the above Lagrangian of the non-symmetric JRM are therefore strongly suppressed and can be neglected. The first non-symmetric contribution in the Lagrangian however reduces to a $\sigma_x\sigma_x$-coupling between the two transmons in the qubit subspace. This term is quadratic in the flux quadratures of the Transmons while our desired $\sigma_z\sigma_z$ coupling term is quartic in the flux quadratures. The inaccuracies in the Josephson energies $\Delta E_i$ may therefore not be larger than 10 percent for typical Transmons,
\begin{equation}
\frac{\varphi_i}{2\varphi_0}=\frac{1}{2}\left(\frac{2E_C}{E_J}\right)^{\frac{1}{4}}(a+a^{\dag})\approx 0.2 (a+a^{\dag})\,.
\end{equation}

\end{document}